\documentclass{revtex4}
\usepackage{graphicx,mathrsfs}
\begin{document}
\title{Extreme waves statistics for Ablowitz-Ladik system}

\author{D.S. Agafontsev$^{(a),(b)}$}
\affiliation{\small \textit{ $^{(a)}$ P. P. Shirshov Institute of Oceanology, 36 Nakhimovsky prosp., Moscow 117218, Russia.\\
$^{(b)}$ Novosibirsk State University, 2 Pirogova, 630090 Novosibirsk, Russia.}}

\begin{abstract}
We examine statistics of waves for the problem of modulation instability development in the framework of discrete integrable Ablowitz-Ladik (AL) system. Modulation instability depends on one free parameter $h$ that has the meaning of the coupling between the nodes on the lattice. For strong coupling $h\ll 1$ the probability density functions (PDFs) for waves amplitudes coincide with that for the continuous classical Nonlinear Schrodinger (NLS) equation; the PDFs for both systems are very close to Rayleigh ones. When the coupling is weak $h\sim 1$, there appear highly localized waves with very large amplitudes, that drastically change the PDFs to significantly non-Rayleigh ones, with so-called "fat tails" when the probability of a large wave occurrence is by several orders of magnitude higher than that predicted by the linear theory. Evolution of amplitudes for such rogue waves with time is similar to that of the Peregrine solution for the classical NLS equation.
\end{abstract}

\maketitle

%------------------------------------------------------------------------------------------------------------------------------------

\textbf{1.} Waves statistics for different nonlinear systems is now one of the most intensively studied topics of nonlinear physics \cite{Hadzievski, Taki1, Bortolozzo, Dudley3, Taki2, Lushnikov, Agafontsev2, Agafontsev3}, especially in connection to rogue waves. Oceanic rogue waves are proved to be dangerous for navigation \cite{Kharif, Dysthe}, while their probability of occurrence is still under discussion. Optical rogue waves \cite{Solli} may damage optical systems, therefore their appearance must be controlled. One of the common scenarios for oceanic and optical rogue waves emergence is realized via nonlinear focusing of waves in the result of the modulation instability development \cite{Solli, Kharif, Dysthe}, described by the classical nonlinear Schrodinger (NLS) equation of focusing type. In this paper we examine waves statistics for the same scenario applied to discrete counterpart of the classical NLS equation, namely to the integrable Ablowitz-Ladik (AL) system \cite{Ablowitz} of focusing 
type:
$$
i\frac{d\Psi_{n}}{dt} +\frac{\Psi_{n+1}-2\Psi_{n}+\Psi_{n-1}}{h^{2}} + \gamma|\Psi_{n}|^{2}\frac{\Psi_{n+1}+\Psi_{n-1}}{2}=0,\quad \Psi_{n}(t=0)=C + \epsilon_{n},
$$
where $n=..., -1, 0, 1, ...$ is node number, $t$ is time, $h>0$ is coupling constant, $\gamma>0$ is nonlinear coefficient, $C$ is level of initial condensate state and $|\epsilon_{n}|\ll|C|$ is a small noise. After the transformations $h^{2}=\tilde{h^{2}}/(\gamma|C|^{2})$, $t=\tilde{t}/(\gamma|C|^{2})$, $\Psi_{n}=C\tilde{\Psi_{n}}e^{i\tilde{t}}$ and $\epsilon_{n}=C\tilde{\epsilon_{n}}e^{i\tilde{t}}$ this problem is reduced to
\begin{equation}\label{Eq01}
i\frac{d\Psi_{n}}{dt} +\frac{\Psi_{n+1}-2\Psi_{n}+\Psi_{n-1}}{h^{2}} - \Psi_{n} + |\Psi_{n}|^{2}\frac{\Psi_{n+1}+\Psi_{n-1}}{2}=0,\quad \Psi_{n}(t=0)=1 + \epsilon_{n},
\end{equation}
where all tilde-signs are omitted. Thus, in contract to the classical NLS equation that can be obtained from Eq. (\ref{Eq01}) after the substitution $x=nh$ in the limit $h\to 0$, problem of modulation instability development for the AL system has one free parameter. 

Let us suppose that the current state of a system consists of multitude of uncorrelated linear waves,
$$
\Psi_{n} = \sum_{k}a_{k}\, e^{i(2\pi kn/M-\omega_{k}t+\phi_{k})},\quad -M/2\le n,k\le M/2-1.
$$
If $a_{k}$ and $\phi_{k}$ are random uncorrelated values and the number of linear waves $M$ is large enough, then under the conditions of central limit theorem real $Re\,\Psi_{n}$ and imaginary $Im\,\Psi_{n}$ parts of field $\Psi_{n}$ are Gaussian-distributed, and the probability to meet amplitude $|\Psi|$ (probability density function, PDF) obeys Rayleigh distribution (see \cite{Agafontsev2}),
$$
PDF(|\Psi|) \sim |\Psi|\exp(-|\Psi|^{2}/2\sigma^{2}).
$$

In this paper we measure PDFs for squared amplitudes that are exponential if the corresponding amplitude PDFs are Rayleigh ones,
$$
PDF(|\Psi|) \sim |\Psi|\exp(-|\Psi|^{2}/2\sigma^{2})\quad \Leftrightarrow \quad PDF(|\Psi|^{2}) \sim \exp(-|\Psi|^{2}/2\sigma^{2}),
$$
and compare the results with exponential dependencies that we call Rayleigh ones for simplicity. In the entire publication we use term "PDF" only in relation to PDFs for (squared) amplitudes of waves. We measure PDFs for entire field $\Psi_{n}$ in contrast to absolute maximums or local maximums PDFs, and use normalization $\int PDF(|\Psi|^{2})\,d|\Psi|^{2}=1$.

In the recent publication \cite{Agafontsev2} it was demonstrated that in case of the modulation instability described by the continuous classical integrable NLS equation the PDFs are generally still very similar to Rayleigh ones, with small time-dependent deviations in the region of medium amplitudes. The aim of the current publication is to extend research made in \cite{Agafontsev2} for the AL equation (\ref{Eq01}), that is also integrable in terms of inverse scattering transformation. For this purpose we take large (10 000 for most of the experiments) ensembles of initial data and examine their evolution with time. Note that in \cite{Hadzievski} there was already done some research in this direction for Salerno system that includes AL system in one of its limits, but the authors actually studied only one model parameter $h=1$. In this publication it will be demonstrated that statistics of waves significantly depends on $h$.\\

\textbf{2.} We solve Eq. (\ref{Eq01}) numerically with the help of Runge-Kutta 4th-order method with time step $\Delta t \le 10^{-4}$ in the box $-M/2 \le n \le M/2-1$ with periodic boundary conditions,
$$
\Psi_{-M/2-1}=\Psi_{M/2-1},\quad \Psi_{M/2}=\Psi_{-M/2}.
$$
Our test simulations confirmed that the PDFs do not depend on the total number of nodes $M$ if the coupling coefficient $h$ is fixed; in our experiments we used number of nodes from $M=128$ to $M=8192$.

As initial data we use condensate solution $\Psi_{n}(t=0)=1+\epsilon_{n}$ perturbed by weak random lattice-homogeneous noise $|\epsilon_{n}|\ll 1$, taken as a superposition of Gaussian-distributed in k-space linear waves,
$$
\epsilon_{n} = \frac{1}{2\pi}\mathscr{F}^{-1}\bigg(A_{0}e^{i\xi_{k}-\kappa_{k}^{2}/\theta^{2}}\bigg)_{n}, \quad -M/2 \le n,k \le M/2-1,
$$
with relatively large dispersion $\theta=5$ (see \cite{Akhmediev}) and arbitrary phases $\xi_{k}$ for each $k$. Here $\kappa_{k}=2\pi k/(h M)$ is wave vector, $\mathscr{F}$ and $\mathscr{F}^{-1}$ are forward and backward discrete Fourier transforms respectively. We use small coefficient $A_{0}=10^{-3}$ corresponding to average noise amplitude $\sqrt{\langle|\epsilon_{n}|^{2}\rangle_{\xi}}\approx 10^{-4}$ (see \cite{Agafontsev3}) in order to ensure that deviations in wave action, energy and momentum inside the ensembles are small. We checked our statistical results obtained with the help of this numerical schema against other lattice-homogeneous statistical distributions of noise and other parameters $A_{0}$ and $\theta$, and also against the size of the ensembles and implementation of other numerical methods, and found no significant difference.\\

\textbf{3.} For the given parameters of initial noise, the nonlinear stage of the modulation instability develops approximately to time shifts $t\sim 10$. The shape of the PDFs highly fluctuate then for the considerable period of time, very similar compared to the classical NLS equation \cite{Agafontsev2}. For small $h$ these fluctuations - excluding Fermi-Pasta-Ulam recurrence points -  diminish to time shifts $t\sim 30$, for $h\sim 1$ - to time shifts $t\sim 100$. After the fluctuations cease, we additionally average the PDFs over time.

\begin{figure}[h] \centering
\includegraphics[width=8cm]{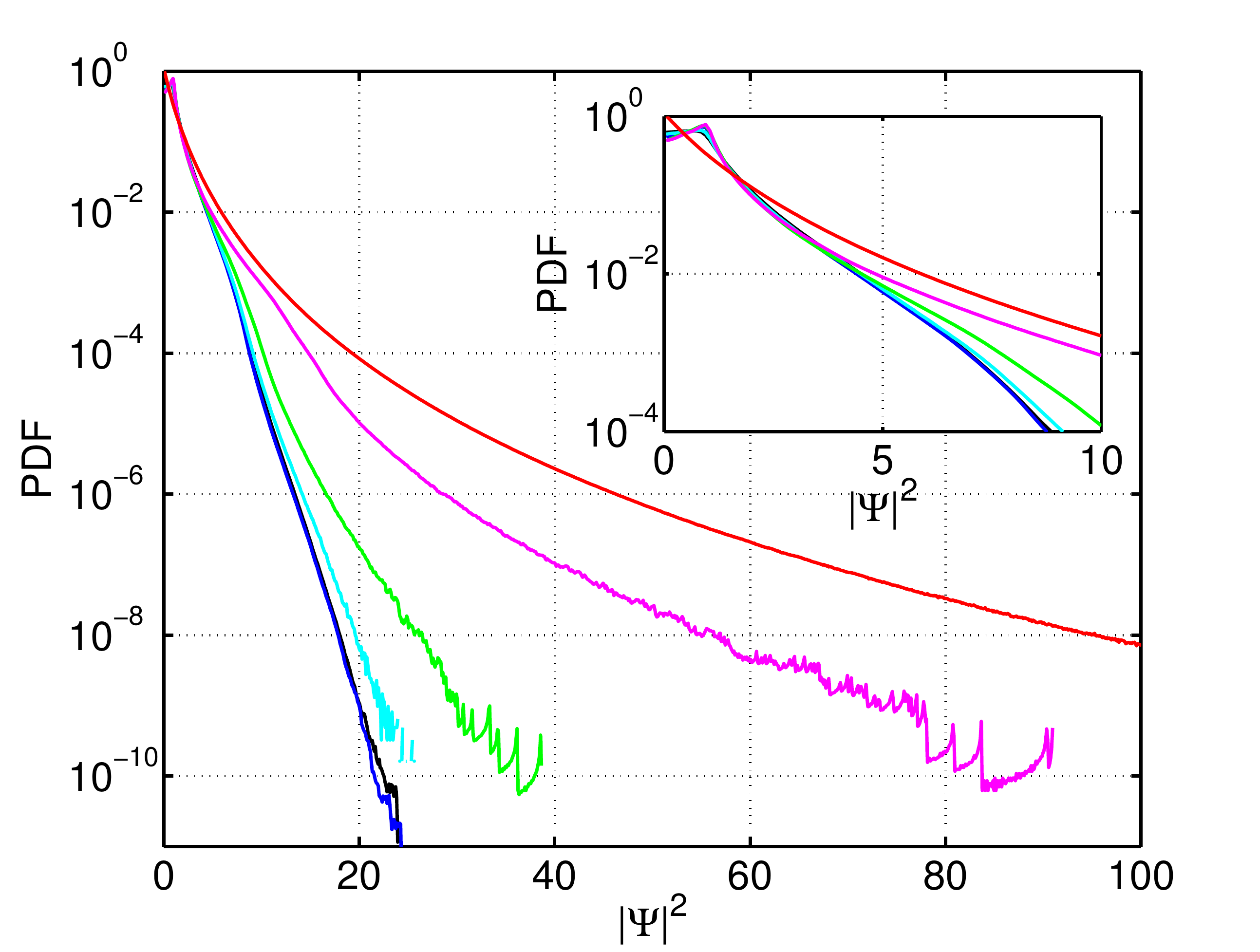}

\caption{\small {\it  (Color on-line) Averaged over ensemble and time $t\in [150, 250]$ squared amplitude PDFs for the continuous classical NLS equation (results from \cite{Agafontsev2}, black) and for Eq. (\ref{Eq01}) with $h=1.2\times 10^{-2}$ (blue), $h=0.20$ (cyan), $h=0.39$ (green), $h=0.79$ (purple) and $h=1$ (red). Inset shows the same lines in the region $|\Psi|^{2}\in[0, 10]$.}}
\label{fig:histogram_h}
\end{figure}

As shown on FIG.~\ref{fig:histogram_h}, for very small coupling coefficients $h$ the resulting PDFs turn out to be exactly the same as for the continuous classical NLS equation: the PDFs for both systems are very close to Rayleigh ones with small but noticeable deviations from Rayleigh shape. These deviations doesn't depend on ensemble size (see \cite{Agafontsev2} for more details) and coincide between the classical NLS equation and the AL model. Significant difference from Rayleigh shape appears starting from $h=0.2$ and then between $h=0.39$ and $h=0.79$ the PDFs become significantly non-Rayleigh ones with severe "fat tails" for large amplitudes, where the probability of occurrence is by several orders of magnitude higher than that predicted by the linear theory.

\begin{figure}[h] \centering
\includegraphics[width=8cm]{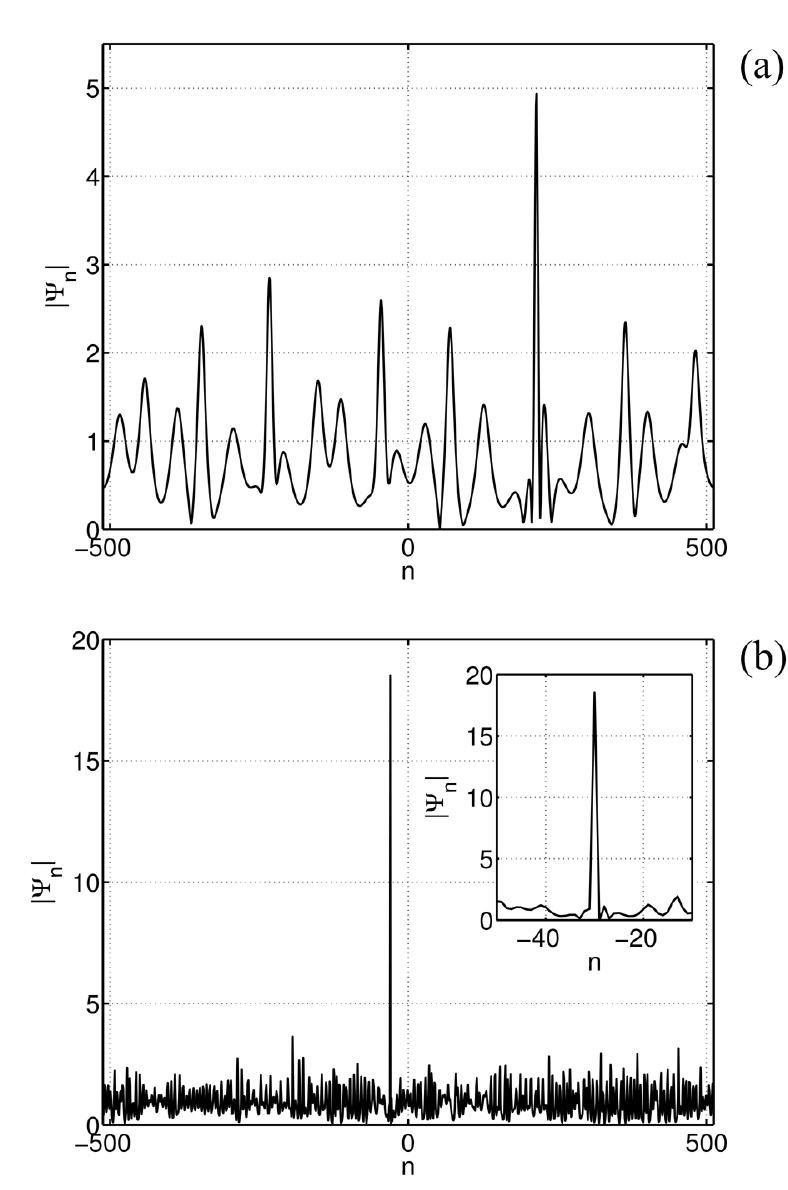}

\caption{\small {\it  Distribution $|\Psi_{n}|$ of a typical large wave event for $h=0.1$ (a) and $h=1$ (b). Inset on graph (b) shows $|\Psi_{n}|$ for $h=1$ in higher node resolution.}}
\label{fig:distribution}
\end{figure}

Typical large wave events are shown on FIG.~\ref{fig:distribution}a,b for $h=0.1$ and $h=1$ cases respectively. For small $h$ the corresponding large wave is very similar to that of the continuous classical NLS equation \cite{Agafontsev2}. In case of $h=1$ there appears highly localized peak occupying one node only. Solution with $h=1$ oscillates with node $n$ significantly faster than with $h=0.1$ because the coupling between the nodes is much weaker.

Effect of energy localization is known for discrete systems for decades (see for example \cite{Turitsyn1, Turitsyn2} and also \cite{Hadzievski}), but so far it related to non-integrable systems with inelastic breather collisions, when higher breathers become higher and smaller breathers become smaller. The same scenario is not valid for the AL system because of the integrability: solitons and breathers collide elastically.

\begin{figure}[h] \centering
\includegraphics[width=8cm]{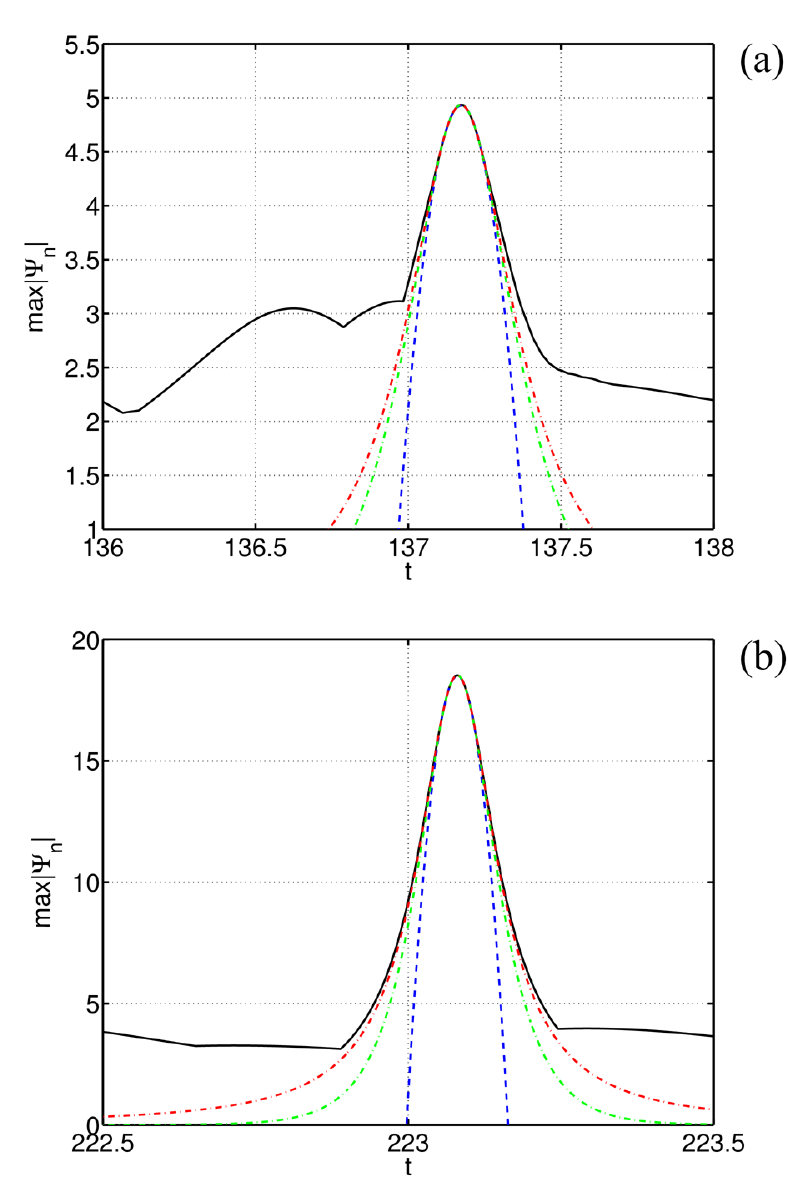}

\caption{\small {\it  (Color on-line) solid black lines - evolution of absolute maximum $\max|\Psi_{n}|$ with time for the same events as on FIG.~\ref{fig:distribution} for $h=0.1$ (a) and $h=1$ (b). Dashed blue lines - fit by parabola $f_{0}(t)=a(t-t_{0})^{2}+c$, dashed-dotted green lines - fit by secant function $f_{1}(t)=A/cosh(\lambda (t-t_{0}))$, dashed-dotted red lines - fit by inverse parabola $f_{2}(t)=1/(a(t-t_{0})^{2}+c)$.}}
\label{fig:fit}
\end{figure}

FIG.~\ref{fig:fit}a,b demonstrate evolution of absolute maximum $\max|\Psi_{n}|$ with time for the same events as on FIG.~\ref{fig:distribution}a,b. In case $h=0.1$ the corresponding large wave event lasts about $\Delta T\approx 0.5$ and approaches in its amplitude to 5, while for $h=1$ it lasts only $\Delta T\approx 0.3$ but it's amplitude hits 18.5. Evolution of absolute maximum was gathered with extremely high temporal resolution of 1 measurement per each $\delta T=10^{-3}$ time shift, that allows us to carefully examine it's dependence on time. We study two hypothesis for the evolution: exponential,
\begin{equation}\label{soliton_collisions}
f_{1}(t)=A/cosh(\lambda (t-t_{0})),  
\end{equation}
and algebraic, similar to the evolution of the Peregrine breather for the classical NLS equation \cite{Akhmediev},
\begin{equation}\label{Peregrine}
f_{2}(t)=1/(a(t-t_{0})^{2}+c).
\end{equation}
For comparison purposes we also include fit by parabola $f_{0}(t)=a(t-t_{0})^{2}+c$. 

As shown on FIG.~\ref{fig:fit}a, it is difficult to separate these two hypothesis for small parameter $h=0.1$: both hypothesis fit pretty well. The situation changes for $h=1$: the algebraic hypothesis \cite{Akhmediev} clearly fits better. We would like to stress, that the large wave events we examined were not specially prepared in any sense, these are random outcomes of the modulation instability that developed from condensate state perturbed by weak random noise.\\

\textbf{4.} In the current publication we examined statistics of waves that appear in the result of the modulation instability development - a common scenario of rogue waves emergence - in the framework of discrete integrable Ablowitz-Ladik system (\ref{Eq01}). We demonstrated that the problem of modulation instability depends on one free parameter $h$ that has the meaning of the coupling coefficient between the nodes on the lattice. In the limit $h\to 0$ the coupling is strong and the AL system transforms into the continuous classical NLS equation with $x=nh$. Our results in this case turn out to be virtually the same as that for the classical NLS equation published in \cite{Agafontsev2}: the PDFs for waves amplitudes are very close to Rayleigh ones with very small but noticeable deviations. These deviations doesn't depend on ensemble size and coincide between the classical NLS equation and the AL system. 

As parameter $h$ increases, the coupling between the nodes decreases, and the PDFs start changing noticeablely from $h=0.2$. Between $h=0.39$ and $h=0.79$ the PDFs become significantly non-Rayleigh ones, resembling the famous L-shape form characteristic to extreme-value processes in some physical systems \cite{Hadzievski, Dudley3, Taki2}, when probability of a large wave appearance is by several orders of magnitude higher than that predicted by the linear theory. 

For small $h$ the extreme waves are smooth pulses with amplitudes up to $\sim 5$ and duration of about $\sim 0.5$. These waves are very similar to that for the classical NLS equation. For weak coupling $h\sim 1$ solutions $\Psi_{n}$ oscillate significantly more frequently with node number $n$, and extreme events represent very high peaks with amplitudes up to $\sim 20$, localized in a very few nodes, and with duration of about $\sim 0.3$. Evolution of their amplitudes with time is very well approximated by the algebraic law, similar to that of the Peregrine solution for the classical NLS equation. For small parameters $h\ll 1$, however, evolution of absolute maximum doesn't allow to separate between exponential or algebraic behavior. We would like to stress that the extreme waves we examined are random outcomes of the modulation instability development and are not initially prepared in any sense.

D. Agafontsev thanks E. Kuznetsov and V. Zakharov for valuable discussions concerning this publication, M. Fedoruk for access to and V. Kalyuzhny for assistance with Novosibirsk Supercomputer Center. This work was done in the framework of Russian Federation Government Grant (contract No. 11.G34.31.0035 with Ministry of Education and Science of RF), and also supported by the program of Presidium of RAS "Fundamental problems of nonlinear dynamics in mathematical and physical sciences", program of support for leading scientific schools of Russian Federation, RFBR grants 12-01-00943-a, 13-01-00261 and also Sergei Badulin RFBR grant 11-05-01114-a.

\end{document}